\documentclass{elsarticle}
\usepackage{graphicx}%
\usepackage{multirow}%
\usepackage{amsmath,amssymb,amsfonts}%
\usepackage{amsthm}%
\usepackage{mathrsfs}%
\usepackage[title]{appendix}%
\usepackage{xcolor}%
\usepackage{textcomp}%
\usepackage{manyfoot}%
\usepackage{booktabs}%
\usepackage{algorithm}%
\usepackage{algorithmicx}%
\usepackage{algpseudocode}%
\usepackage{listings}%
\usepackage{verbatim}
\usepackage{enumitem}
 \usepackage{comment}
 \usepackage{amsmath}
 \usepackage{booktabs}
\usepackage{amssymb}

\journal{}
\begin{document}

\begin{frontmatter}

\title{Renewal Bias  in Parallel Simulations of Absorbing Processes}

\author[GMU]{Sandro M. Reia}
%\cortext[mycorrespondingauthor]{Corresponding author}
%\ead{smarti71@gmu.edu }

\author[mymainaddress]{Jos\'e F.  Fontanari}
%\ead{fontanari@ifsc.usp.br}

\address[GMU]{Geography and Geoinformation Science, College of Science, George Mason University, Fairfax, VA, USA}

\address[mymainaddress]{Instituto de F\'{\i}sica de S\~ao Carlos,  Universidade de S\~ao Paulo,  13566-590 S\~ao Carlos,  S\~ao Paulo,  Brazil}

\begin{abstract}
Asynchronous parallel simulations of stochastic processes with absorbing states can exhibit systematic sampling bias when trajectory durations depend on their eventual absorbing outcomes. Workers that continuously restart upon absorption preferentially record rapidly terminating trajectories within a fixed computational budget, distorting the observed ensemble statistics. We investigate this effect using a minimal renewal model with two absorbing states and state-dependent exponential durations. Renewal theory yields exact expressions for the expected number and composition of completed trajectories, showing that the resulting bias decreases algebraically with execution time but persists as the number of parallel workers increases at fixed budget. We compare continuous sampling with First-Trajectory Filtering, which retains only the initial trajectory attempted by each worker. Although finite execution deadlines introduce residual censoring bias, this bias decreases exponentially with the computational budget. An analytical mean-squared error calculation reveals budget-dependent crossover worker allocations beyond which First-Trajectory Filtering becomes more accurate despite collecting fewer trajectories. We also examine the limitations of statistical inference under severe censoring when the absorbing-state probabilities and characteristic durations are unknown. Finally, simulations of the Axelrod model near its nonequilibrium transition demonstrate the preferential sampling of rapidly absorbing fragmented configurations, illustrating the relevance of the renewal mechanism in a spatially interacting system. Our results show that increasing computational throughput does not necessarily improve statistical accuracy and highlight the importance of accounting for trajectory-duration heterogeneity in parallel simulations of absorbing processes. 
\end{abstract}

%\keywords{synchronization, collective motion, correlated random-walk, spatial accessibility}

%%\pacs[JEL Classification]{D8, H51}

%%\pacs[MSC Classification]{35A01, 65L10, 65L12, 65L20, 65L70}

\end{frontmatter}

\section{Introduction}\label{sec:intro}

Stochastic dynamical systems with absorbing states arise in a broad range of problems, including cultural dissemination~\cite{Axelrod_1997}, voter and consensus dynamics~\cite{Clifford_1973,Holley_1975}, mutant fixation in population genetics~\cite{Moran_1958}, and stochastic epidemic outbreaks~\cite{Bailey_1975}. In finite systems that reach absorption, independent realizations may terminate in qualitatively different configurations, often after widely varying transient times. When the duration of a trajectory is correlated with its eventual absorbing state, the procedure used to collect completed realizations can influence the observed ensemble statistics.

This issue is particularly relevant to high-performance computing (HPC), where large ensembles of independent stochastic simulations are commonly distributed across parallel workers operating under fixed wall-clock limits~\cite{Voter_1998,Kroese_2011}. An efficient execution strategy is to restart a worker immediately after each trajectory reaches absorption, thereby maximizing the number of completed realizations within the allocated computational budget. However, this continuous sampling procedure does not necessarily produce a representative ensemble at finite execution times. Rapidly absorbing trajectories can be completed and recorded more frequently, whereas trajectories with longer transients remain unfinished when the execution deadline is reached. Consequently, the distribution of completed outcomes may differ systematically from that obtained by independently sampling initial conditions and following every trajectory to absorption.

The resulting distortion involves two closely related aspects of the sampling protocol. First, repeated worker renewal allows rapidly absorbing outcomes to contribute disproportionately to the accumulated dataset during finite execution windows. Second, the hard execution deadline introduces right-censoring, leaving the eventual absorbing states of incomplete trajectories unobserved~\cite{Klein_2003}. These effects are especially important when the characteristic absorption times differ substantially between competing absorbing configurations. Increasing the number of parallel workers reduces statistical fluctuations, but does not necessarily eliminate the systematic bias associated with a fixed execution budget.

To investigate this problem, we introduce a minimal renewal model in which independent trajectories terminate in one of two absorbing states with prescribed probabilities and state-dependent exponential durations. The model isolates the statistical consequences of heterogeneous absorption times without introducing the microscopic complexity of a particular interacting system. Its exponential structure permits an analytical treatment using renewal theory, yielding exact expressions for the expected number of completed trajectories and their absorbing-state composition. We show that continuous sampling exhibits a finite-budget bias that decreases algebraically with execution time, while remaining nonzero in the limit of infinitely many workers at fixed budget.

We also investigate First-Trajectory Filtering, a sampling strategy that retains only the initial trajectory attempted by each worker. By eliminating repeated renewals from the analyzed sample, this procedure substantially reduces the systematic distortion. Nevertheless, when initial trajectories are subject to a finite deadline, their state-dependent completion probabilities introduce a residual selection bias. For the exponential model, this bias decreases exponentially with the computational budget. We derive the corresponding mean-squared error (MSE) and compare the statistical efficiency of First-Trajectory Filtering with that of continuous sampling under identical computational resource allocations. The resulting competition between sampling variance and systematic bias produces budget-dependent crossover scales in the number of parallel workers.

A further complication arises when the absorbing-state probabilities and conditional duration distributions are unknown, as is generally the case in simulations of complex stochastic systems. Under severe right-censoring, the completed trajectories may provide insufficient information for reliable estimation of the unobserved outcomes. We therefore examine the limitations of statistical inference at short execution times and distinguish the availability of formal bias corrections from their practical efficiency when the required model parameters must themselves be estimated.

As a physical application, we consider the Axelrod model of cultural dissemination near its nonequilibrium transition~\cite{Castellano_2000,Klemm_2003,Reia_2016,Campos_2025,Reia_2026}. In this system, two distinct classes of absorbing configurations exhibit markedly different characteristic absorption times, providing a natural motivation for the binary structure of our minimal model. We show that continuous sampling preferentially represents rapidly absorbing fragmented configurations, leading to a systematic displacement of the observed domain density relative to independent reference trajectories followed to absorption. The Axelrod simulations thus illustrate the renewal-sampling mechanism in a nontrivial spatial system without requiring the exponential model to reproduce its detailed microscopic dynamics.

The remainder of the paper is organized as follows. Section~\ref{sec:mapping} relates wall-clock execution limits to effective computational budgets. Section~\ref{sec:model} introduces the exponential renewal model and the sampling protocols. In Sec.~\ref{sec:results}, we derive the analytical predictions for continuous sampling and First-Trajectory Filtering, examine their MSE trade-offs, and discuss observability limitations under severe censoring. Section~\ref{sec:axelrod} illustrates the sampling mechanism in the Axelrod model. Finally, Sec.~\ref{sec:discussion} summarizes the implications for asynchronous parallel simulations of absorbing processes.

\section{Mapping Wall-Clock Time to Worker Iteration Budgets}
\label{sec:mapping}

To relate physical computational constraints to simulation dynamics, we distinguish between wall-clock time and algorithmic time. Let $t_{\text{wall}}$ denote the physical wall-clock time limit allocated to each worker by the high-performance computing (HPC) scheduler, and let $\omega$ denote the processor execution rate, measured in elementary state updates (or Monte Carlo steps) per unit wall-clock time. Assuming a constant execution rate, the effective computational budget available to each worker is
\begin{equation}
    T_{\text{budget}} = \omega t_{\text{wall}},
\end{equation}
where $T_{\text{budget}}$ is measured in units of state updates.

The distinction is relevant because the computational cost of an elementary update generally depends on the simulated system. For example, increasing the number of agents or lattice sites may reduce the execution rate $\omega$, thereby decreasing the number of updates that can be performed within a fixed wall-clock allocation. Consequently, a fixed physical time limit may correspond to different effective computational budgets for different system sizes, potentially increasing the fraction of trajectories that remain incomplete when the allocation expires.

In the following, we measure trajectory durations directly in effective com\-pu\-ta\-tion\-al-time units and represent the wall-clock constraint through $T_{\text{budget}}$. This allows us to isolate the statistical consequences of finite execution budgets without explicitly modeling processor-level details.

\section{Model Specification and Scale Invariance}
\label{sec:model}

We consider a cluster of $M$ independent parallel workers operating asynchronously. Each worker generates a sequence of independent stochastic trajectories, each terminating in one of two absorbing states $X\in\{0,1\}$. The absorbing-state probabilities are
\begin{equation}
    P(X=0)=a,\qquad P(X=1)=b=1-a.
\end{equation}
Although the absorbing state is drawn at the beginning of each trajectory in the minimal model, it represents an outcome that becomes observable only upon absorption.

The absorption time $T>0$ is drawn from a state-dependent exponential distribution,
\begin{align}
    f(t|X=1)&=\frac{1}{\tau_1}
    \exp\left(-\frac{t}{\tau_1}\right),
    \label{eq:fast}\\
    f(t|X=0)&=\frac{1}{\tau_0}
    \exp\left(-\frac{t}{\tau_0}\right),
    \label{eq:slow}
\end{align}
where $\tau_1$ and $\tau_0$ are the corresponding mean absorption times. We assume $\tau_1<\tau_0$, so trajectories ending in state $X=1$ typically reach absorption faster than those ending in state $X=0$. Since exponential absorption times have unbounded support, for any finite execution budget, there is a nonzero probability that a trajectory remains incomplete.

The unconditional absorption-time density is the exponential mixture
\begin{equation}
    f(t)=
    \frac{b}{\tau_1}\exp\left(-\frac{t}{\tau_1}\right)
    +\frac{1-b}{\tau_0}\exp\left(-\frac{t}{\tau_0}\right),
    \label{eq:mixture}
\end{equation}
with mean
\begin{equation}
    \bar{\tau}\equiv\mathbb{E}[T]
    =b\tau_1+(1-b)\tau_0.
    \label{eq:mean_duration}
\end{equation}
Unless $\tau_1=\tau_0$, the mixture is not itself exponentially distributed, and the sequence of trajectory completions constitutes a renewal process rather than a Poisson process.

Throughout the numerical analysis, we choose $\tau_1=1/6$ and $\tau_0=5/6$, corresponding to a fivefold difference in mean absorption times. These values set the relative timescales of the two absorbing outcomes, while the prior probability $b$ controls their relative frequencies.

Data collection proceeds independently across workers. Each worker $m\in\{1,\ldots,M\}$ maintains an accumulated computational-time counter $t_{\text{acc}}$, initially set to zero. For each new trajectory, an absorbing state $X_k$ is independently drawn from the Bernoulli distribution with parameter $b$, followed by an absorption time $T_k$ drawn from the corresponding conditional exponential distribution.

If $t_{\text{acc}}+T_k\leq T_{\text{budget}}$, the trajectory is recorded as completed, the counter is updated according to $t_{\text{acc}}\leftarrow t_{\text{acc}}+T_k$, and a new trajectory is immediately initiated. Otherwise, the trajectory is right-censored at the execution deadline. Its elapsed duration is $\Delta t=T_{\text{budget}}-t_{\text{acc}}$, while its eventual absorbing state remains unobserved.

Let $N$ denote the aggregate number of completed trajectories collected across the $M$ workers. The cumulative observed mean is
\begin{equation}
    \langle X\rangle_{\text{obs}}(N)
    =\frac{1}{N}\sum_{i=1}^{N}X_i,
    \label{eq:observed_mean}
\end{equation}
where the sum includes only completed trajectories. Because the absorption-time distribution depends on the absorbing state, the set of trajectories completed within a finite budget need not reproduce the underlying prior probability $b$.

We also consider First-Trajectory Filtering, in which only the initial trajectory attempted by each worker is retained for statistical analysis. If an initial trajectory is incomplete at the execution deadline, its absorbing state is unavailable and it is excluded from the observed sample. Thus, even this filtered estimator may exhibit selection bias under a finite computational budget, although the bias decreases as the probability of completing an initial trajectory approaches unity.

A useful dimensionless control parameter is the ratio of the available computational budget to the unconditional mean absorption time,
\begin{equation}
    \gamma\equiv
    \frac{T_{\text{budget}}}{\bar{\tau}}
    =\frac{\omega t_{\text{wall}}}
    {b\tau_1+(1-b)\tau_0}.
    \label{eq:gamma}
\end{equation}
Introducing the normalized absorption time $y=T/\bar{\tau}$, the conditional densities become
\begin{equation}
    f(y|X=x)=
    \frac{\bar{\tau}}{\tau_x}
    \exp\left(-\frac{\bar{\tau}}{\tau_x}y\right),
    \qquad x\in\{0,1\}.
    \label{eq:normalized_exp}
\end{equation}
The condition for completing $k$ consecutive trajectories is then
\begin{equation}
    \sum_{i=1}^{k}y_i\leq\gamma.
\end{equation}
Consequently, the sampling statistics depend on the dimensionless budget $\gamma$, the prior probability $b$, and the ratio $\tau_0/\tau_1$, rather than on the absolute absorption-time scale. A common rescaling of $\tau_0$, $\tau_1$, and $T_{\text{budget}}$ leaves the normalized process unchanged. Table~\ref{tab:parameters} summarizes the model parameters and their physical meanings, including the computational quantities introduced in Sec.~\ref{sec:mapping}.

\begin{table}[htbp]
\centering
\caption{Control parameters of the asynchronous sampling model.}
\label{tab:parameters}
\resizebox{\linewidth}{!}{%
\begin{tabular}{ccc}
\toprule
\textbf{Parameter} & \textbf{Domain} & \textbf{Physical meaning / description}\\
\midrule
$a,b$ & $[0,1]$, $a+b=1$ & Prior probabilities of absorbing states $X=0,1$.\\
$\tau_1,\tau_0$ & $\mathbb{R}^{+}$ & Mean absorption times for states $X=1,0$.\\
$\bar{\tau}$ & $\mathbb{R}^{+}$ & Unconditional mean absorption time.\\
$t_{\text{wall}}$ & $\mathbb{R}^{+}$ & Physical wall-clock allocation per worker.\\
$\omega$ & $\mathbb{R}^{+}$ & State updates per unit wall-clock time.\\
$T_{\text{budget}}$ & $\mathbb{R}^{+}$ & Effective computational budget per worker.\\
$\gamma$ & $\mathbb{R}^{+}$ & Dimensionless budget ratio $T_{\text{budget}}/\bar{\tau}$.\\
$M$ & $\mathbb{Z}^{+}$ & Number of independent parallel workers.\\
\bottomrule
\end{tabular}%
}
\end{table}

The dimensionless budget controls the importance of finite-time sampling effects. For $\gamma\ll1$, only a small fraction of trajectories reach absorption, and completed samples are strongly enriched in the faster state $X=1$. For $\gamma$ of order unity, workers complete different numbers of renewal cycles, and the observed sample mean may exhibit pronounced transient deviations from the underlying prior. Finally, for $\gamma\gg1$, each worker completes on average approximately $\gamma$ trajectories, and the influence of the incomplete boundary trajectory becomes progressively smaller. In this large-budget regime, the cumulative observed mean approaches the true prior $b$.

Importantly, increasing the number of workers $M$ at fixed $\gamma$ reduces statistical fluctuations but does not, in general, eliminate finite-budget sampling bias. By contrast, increasing $\gamma$ allows each worker to complete more renewal cycles and progressively suppresses the bias. This distinction between worker allocation and execution depth will be central to the analysis of estimator accuracy and computational efficiency in the following section.

\section{Results}\label{sec:results}

We investigate how finite computational budgets affect the statistical accuracy of asynchronous sampling in the exponential renewal model. Our analysis focuses on two complementary aspects: the evolution of the observed absorbing-state probability under continuous sampling and the mean-squared error of estimators constructed from completed trajectories. Whenever possible, we compare numerical simulations with analytical predictions obtained directly from the renewal process, without adjustable parameters. Particular attention is given to the distinction between increasing the number of parallel workers and extending the execution time available to each worker.

\subsection{Macroscopic Estimator Dynamics and Asymptotic Convergence}

\begin{figure}[htbp]
    \centering
    \includegraphics[width=1.0\linewidth]{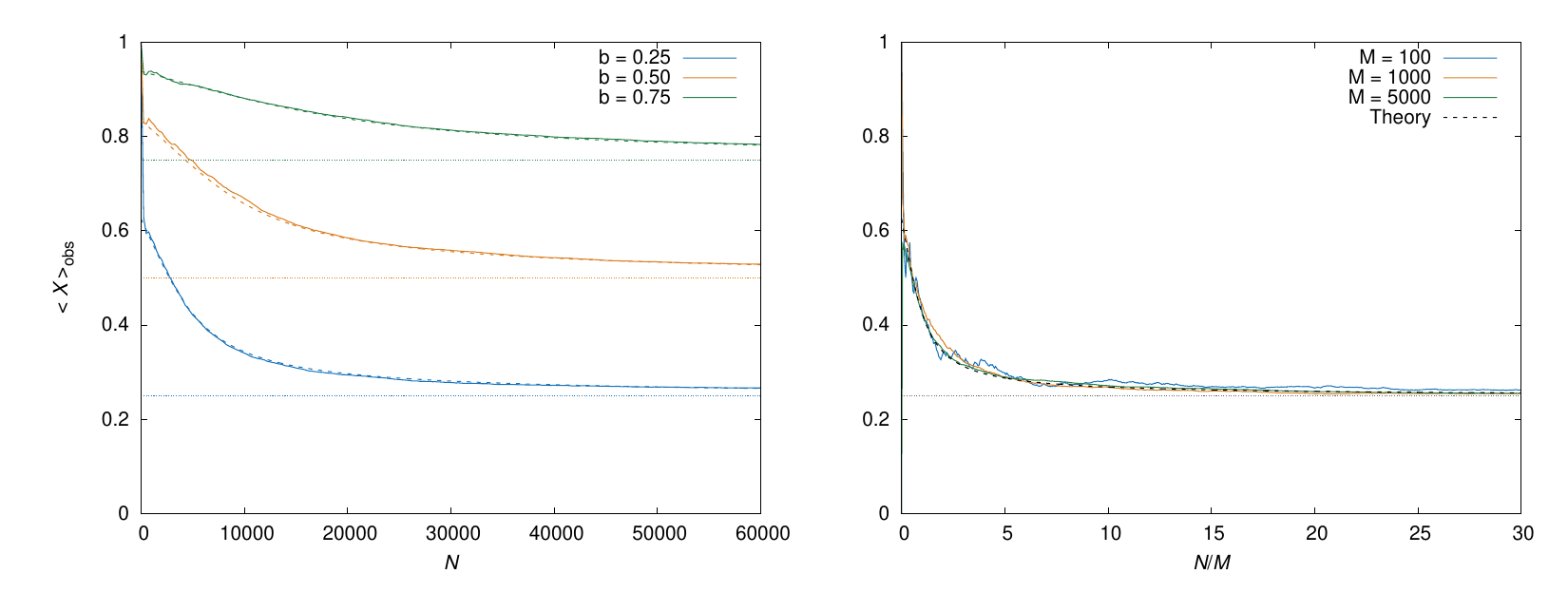}
    \caption{
    Evolution of the cumulative observed mean
    $\langle X\rangle_{\mathrm{obs}}$ in the exponential
    renewal model.
    Left panel: Observed mean versus the total number
    $N$ of completed trajectories for prior probabilities
    $b=0.25$, $0.50$, and $0.75$, with $M=5000$
    parallel workers. Horizontal dotted lines indicate
    the corresponding true prior probabilities.
    Right panel: Observed mean versus the average number
    $N/M$ of completed trajectories per worker for
    $M=100$, $1000$, and $5000$, with $b=0.25$.
    Colored solid lines represent numerical simulations,
    while black dashed lines show the analytical
    large-$M$ predictions obtained from
    Eqs.~\eqref{eq:renewal_mean}--\eqref{eq:observed_exact},
    without adjustable parameters.
    The conditional mean absorption times are
    $\tau_1=1/6$ and $\tau_0=5/6$, and the maximum
    normalized execution budget is $\gamma_{\max}=30$.
    The left panel displays the initial portion of
    the simulated trajectories.
    }
    \label{fig:mean_x_vs_b}
\end{figure}

The left panel of Fig.~\ref{fig:mean_x_vs_b} shows the evolution of the cumulative observed mean $\langle X\rangle_{\mathrm{obs}}$ as a function of the total number $N$ of completed trajectories for three underlying probabilities, $b=0.25$, $0.50$, and $0.75$. In all cases, the initially observed fraction of trajectories ending in state $X=1$ substantially exceeds its true probability $b$. This overrepresentation arises because trajectories reaching state $X=1$ have a shorter mean absorption time and are therefore more likely to contribute to the sample during the early stages of execution. As the computational budget increases, the observed mean gradually approaches the corresponding prior, indicated by the horizontal dotted lines.

The transient evolution can be calculated analytically using renewal theory~\cite{Cox_1962,Ross_1996}. Let $K(t)$ denote the number of trajectories completed by a single worker up to computational time $t$, and let $R(t)$ denote the number of those trajectories ending in state $X=1$. We introduce the mean completion counts
\begin{equation}
    m(t)=\mathbb{E}[K(t)],
    \qquad
    r(t)=\mathbb{E}[R(t)].
\end{equation}
Because successive trajectories are independent, their completion times form a renewal process with waiting-time density $f(t)$ given by Eq.~\eqref{eq:mixture}.

Writing $\lambda_1=1/\tau_1$ and $\lambda_0=1/\tau_0$, the Laplace transform of the waiting-time density is
\begin{equation}
    \widetilde f(s)=
    \frac{b\lambda_1}{s+\lambda_1}
    +\frac{(1-b)\lambda_0}{s+\lambda_0}.
\end{equation}
The renewal equation for the expected number of completed trajectories gives
\begin{equation}
    \widetilde m(s)=
    \frac{\widetilde f(s)}
    {s[1-\widetilde f(s)]}.
\end{equation}
Similarly, the expected number of completed trajectories ending in state $X=1$ satisfies
\begin{equation}
    \widetilde r(s)=
    \frac{b\lambda_1}
    {s(s+\lambda_1)[1-\widetilde f(s)]}.
\end{equation}

Defining
\begin{equation}
    c=b\lambda_0+(1-b)\lambda_1,
\end{equation}
the inverse Laplace transforms yield
\begin{equation}
    m(t)=\frac{t}{\bar\tau}
    +\frac{b(1-b)(\lambda_1-\lambda_0)^2}{c^2}
    \left(1-e^{-ct}\right)
    \label{eq:renewal_mean}
\end{equation}
and
\begin{equation}
    r(t)=bm(t)
    +\frac{b(1-b)(\lambda_1-\lambda_0)}{c}
    \left(1-e^{-ct}\right).
    \label{eq:renewal_reward}
\end{equation}
These expressions are exact for the exponential mixture and describe the entire transient evolution of the mean completion counts.

For $M$ independent workers, the total number of completed trajectories and the number ending in state $X=1$ have expectations $Mm(t)$ and $Mr(t)$, respectively. Consequently, in the large-$M$ limit, the cumulative observed mean converges to
\begin{equation}
    \langle X\rangle_{\mathrm{th}}(t)
    =\frac{r(t)}{m(t)}
    =b+
    \frac{
    b(1-b)(\lambda_1-\lambda_0)
    \left(1-e^{-ct}\right)
    }{
    c\,m(t)
    }.
    \label{eq:observed_exact}
\end{equation}
Here the ratio of expectations provides the deterministic large-$M$ prediction; for finite $M$, the expectation of the observed ratio is not generally identical to $r(t)/m(t)$.

Since $\tau_1<\tau_0$, we have $\lambda_1>\lambda_0$, and Eq.~\eqref{eq:observed_exact} predicts
$\langle X\rangle_{\mathrm{th}}(t)>b$ for every finite $t>0$. Thus, the overrepresentation of the faster-absorbing state is an intrinsic finite-time property of the sampling process.

In the short-time limit, Eq.~\eqref{eq:observed_exact} reduces to
\begin{equation}
    \lim_{t\to0^+}
    \langle X\rangle_{\mathrm{th}}(t)
    =
    \frac{b\lambda_1}
    {b\lambda_1+(1-b)\lambda_0}.
    \label{eq:initial_bias}
\end{equation}
This result has a simple interpretation: at very short times, the probability that a trajectory has completed is proportional to its absorption rate. For the parameters $\tau_1=1/6$, $\tau_0=5/6$, and $b=0.25$, the initial observed probability is $0.625$, considerably larger than the true value $0.25$.

At long times, the exponential terms in Eqs.~\eqref{eq:renewal_mean} and \eqref{eq:renewal_reward} become negligible. Using $m(t)\sim t/\bar\tau$, we obtain
\begin{equation}
    \langle X\rangle_{\mathrm{th}}(t)-b
    \sim
    \frac{b(1-b)(\tau_0-\tau_1)}{t}.
    \label{eq:asymptotic_time}
\end{equation}
The systematic bias therefore decreases inversely with the execution time, independently of the number of workers.

To express this result in terms of the accumulated sample count, we use the large-$M$ relation
\begin{equation}
    N_{\mathrm{th}}(t)=Mm(t)
    \sim\frac{Mt}{\bar\tau}.
\end{equation}
Substituting into Eq.~\eqref{eq:asymptotic_time} gives
\begin{equation}
    \langle X\rangle_{\mathrm{th}}-b
    \sim
    \frac{b(1-b)(\tau_0-\tau_1)}{\bar\tau}
    \frac{M}{N_{\mathrm{th}}}.
    \label{eq:1overN_decay}
\end{equation}
Thus, the inverse-sample-count decay observed in Fig.~\ref{fig:mean_x_vs_b} is not an empirical fitting law: its amplitude follows directly from the renewal model. Importantly, the coefficient of $1/N_{\mathrm{th}}$ is proportional to $M$. Increasing the number of workers at fixed execution time increases the total sample count without eliminating the systematic bias.

The analytical predictions shown in Fig.~\ref{fig:mean_x_vs_b} are obtained by plotting Eq.~\eqref{eq:observed_exact} parametrically against $N_{\mathrm{th}}(t)=Mm(t)$. The close agreement with the simulations confirms that the transient overrepresentation of state $X=1$ is quantitatively captured by the renewal description, without adjustable parameters.

%----------------------------------------
%----------------------------------------

The right panel of Fig.~\ref{fig:mean_x_vs_b} provides a complementary test of the role of parallelism. For a given prior $b$, the theoretical observed mean depends on computational time through the single-worker renewal functions $m(t)$ and $r(t)$. Eliminating $t$ between these functions produces a relation between $\langle X\rangle_{\mathrm{th}}$ and the average number of completed trajectories per worker,
\begin{equation}
    k\equiv\frac{N_{\mathrm{th}}}{M}=m(t),
\end{equation}
which is independent of $M$. Consequently, simulations with different cluster sizes are expected to approach the same master curve when plotted against $N/M$.

In the large-$k$ regime, Eq.~\eqref{eq:1overN_decay} becomes
\begin{equation}
    \langle X\rangle_{\mathrm{th}}-b
    \sim\frac{A(b)}{k},
    \qquad
    A(b)=
    \frac{b(1-b)(\tau_0-\tau_1)}{\bar\tau}.
    \label{eq:data_collapse}
\end{equation}
For $b=0.25$ and the absorption times used here, $A(b)=3/16$. The asymptotic bias is therefore approximately $3/(16k)$, providing a quantitative prediction for the late-time behavior in the right panel.

The numerical results confirm the predicted data collapse. The curves for $M=1000$ and $M=5000$ closely follow the analytical prediction, while the smaller cluster with $M=100$ exhibits larger statistical fluctuations. These fluctuations decrease with increasing $M$, whereas the systematic finite-budget bias remains unchanged in the large-$M$ limit.

More precisely, at a fixed computational time $t$, the completed trajectory counts from different workers are independent. Therefore, the fluctuations of their aggregate ratio decrease as $M^{-1/2}$, provided the relevant moments are finite. The variance coefficient, however, depends on the joint fluctuations of the numbers of completed trajectories and their absorbing states; it is not generally given by the Bernoulli expression $b(1-b)/N$, because trajectory duration and absorbing state are correlated.

This distinction has an important operational consequence. Increasing the number of parallel workers improves statistical precision by reducing fluctuations around the finite-time mean, but does not remove the systematic overrepresentation of the faster-absorbing state. Reducing this bias requires increasing the execution time available to each worker, or modifying the sampling procedure. The competition between these two strategies motivates the comparison of continuous sampling and First-Trajectory Filtering developed below.

\subsection{First-Trajectory Filtering: Residual Censoring Bias and Statistical Precision}
\label{sec:first_trajectory}

The preceding analysis shows that the systematic bias of continuous sampling decreases algebraically with the computational budget. An alternative strategy is First-Trajectory Filtering, in which only the initial trajectory attempted by each worker is retained for statistical analysis. Subsequent renewal cycles are discarded, regardless of whether they reach absorption. This procedure eliminates the repeated-sampling mechanism responsible for the transient renewal bias, but it does not remove the selection bias associated with incomplete initial trajectories.

For the exponential model, the residual bias can be calculated explicitly. Let $B=T_{\mathrm{budget}}=\gamma\bar\tau$ denote the computational budget. The probability that an initial trajectory reaches absorption before the deadline, conditional on its absorbing state, is
\begin{equation}
    q_x(B)=P(T\leq B\mid X=x)
    =1-\exp(-B/\tau_x),
    \qquad x\in\{0,1\}.
    \label{eq:first_completion}
\end{equation}
The unconditional completion probability is therefore
\begin{equation}
    q(B)=bq_1(B)+(1-b)q_0(B).
    \label{eq:first_q}
\end{equation}
Because $\tau_1<\tau_0$, we have $q_1(B)>q_0(B)$ for every finite $B>0$. Thus, trajectories ending in state $X=1$ are overrepresented among completed initial trajectories.

The probability of observing state $X=1$, conditional on completion, follows from Bayes' rule:
\begin{equation}
    p(B)\equiv P(X=1\mid T\leq B)
    =\frac{bq_1(B)}{q(B)}.
    \label{eq:first_probability}
\end{equation}
The resulting selection bias is
\begin{equation}
    p(B)-b=
    \frac{
    b(1-b)\left[
    e^{-B/\tau_0}-e^{-B/\tau_1}
    \right]
    }{
    1-be^{-B/\tau_1}-(1-b)e^{-B/\tau_0}
    }.
    \label{eq:first_bias}
\end{equation}
This expression is positive for every finite budget and vanishes exponentially as $B$ increases. In particular, for $0<b<1$ and $\tau_1<\tau_0$,
\begin{equation}
    p(B)-b
    \sim b(1-b)e^{-B/\tau_0},
    \qquad B\to\infty.
    \label{eq:first_bias_asymptotic}
\end{equation}
This contrasts with the algebraic decay of the continuous-sampling bias obtained in Eq.~\eqref{eq:asymptotic_time}. First-Trajectory Filtering therefore suppresses the systematic error much more rapidly as the execution budget increases, although it is not exactly unbiased at any finite deadline.

To quantify the statistical accuracy of First-Trajectory Filtering, let $J$ denote the number of workers whose initial trajectories are completed within the computational budget. Since workers are independent,
\begin{equation}
    J\sim\mathrm{Binomial}(M,q(B)).
\end{equation}
For $J>0$, the estimator of the absorbing-state probability $b$ is defined as
\begin{equation}
    \hat b_{\mathrm{first}}
    =\frac{1}{J}
    \sum_{m=1}^{M}
    X_{m,1}\,
    \mathbb{I}(T_{m,1}\leq B),
    \label{eq:b_first}
\end{equation}
where $\mathbb{I}$ denotes the indicator function and the subscript $1$ identifies the initial trajectory of each worker.

The accuracy of an estimator $\hat b$ of the true absorbing-state probability $b$ can be characterized by its mean-squared error (MSE),
\begin{equation}
    \mathrm{MSE}(\hat b)
    =\mathbb{E}\left[(\hat b-b)^2\right]
    =\left[\mathbb{E}(\hat b)-b\right]^2
    +\mathrm{Var}(\hat b).
    \label{eq:mse_definition}
\end{equation}
Here, $\hat b$ denotes a generic estimator, such as $\hat b_{\mathrm{first}}$ or the continuous-sampling estimator. The first term measures the squared systematic bias, while the second quantifies statistical fluctuations. This decomposition is particularly useful because increasing the number of workers reduces sampling fluctuations without necessarily eliminating the bias associated with a finite execution budget.

Conditional on $J=j>0$, the observed absorbing states are independent Bernoulli variables with success probability $p(B)$. Consequently,
\begin{equation}
    \mathbb{E}[\hat b_{\mathrm{first}}\mid J=j]
    =p(B),
\end{equation}
and
\begin{equation}
    \mathrm{Var}(\hat b_{\mathrm{first}}\mid J=j)
    =\frac{p(B)[1-p(B)]}{j}.
\end{equation}

Thus, conditional on at least one completed trajectory, the MSE is
\begin{equation}
\begin{split}
    \mathrm{MSE}_{\mathrm{first}}^{(+)}
    ={}&[p(B)-b]^2\\
    &+p(B)[1-p(B)]
    \mathbb{E}\left[\frac{1}{J}\middle|J>0\right].
\end{split}
    \label{eq:first_mse_conditional}
\end{equation}
The first term represents the squared selection bias, whereas the second accounts for statistical fluctuations in the completed sample, including fluctuations in the number $J$ of observed trajectories.

For direct comparison with our Monte Carlo simulations, we adopt the convention $\hat b_{\mathrm{first}}=0$ when $J=0$. Since
$P(J=0)=[1-q(B)]^M$, the corresponding unconditional MSE is
\begin{equation}
\begin{split}
    \mathrm{MSE}_{\mathrm{first}}
    ={}&[1-(1-q)^M](p-b)^2\\
    &+p(1-p)
    \sum_{j=1}^{M}
    \binom{M}{j}
    \frac{q^j(1-q)^{M-j}}{j}\\
    &+b^2(1-q)^M,
\end{split}
    \label{eq:first_mse_exact}
\end{equation}
where $p=p(B)$ and $q=q(B)$. Equation~\eqref{eq:first_mse_exact} is exact for arbitrary $M$ and $B$ and provides the analytical prediction used in our numerical comparison. The selection bias \(p(B)-b\) refers to samples containing at least one completed trajectory; under the zero-estimate convention, the unconditional finite-\(M\) bias also includes the probability that no initial trajectory completes.

For sufficiently large $M$, the probability of observing no completed trajectories becomes negligible, and $J$ is concentrated around its mean $Mq(B)$. Equation~\eqref{eq:first_mse_exact} then simplifies to
\begin{equation}
    \mathrm{MSE}_{\mathrm{first}}
    \simeq
    [p(B)-b]^2+
    \frac{p(B)[1-p(B)]}{Mq(B)}.
    \label{eq:first_mse_approx}
\end{equation}
The approximation makes the competition between bias and variance particularly transparent. At fixed $B$, the variance decreases approximately as $1/M$, while the squared bias remains unchanged. Hence,
\begin{equation}
    \lim_{M\to\infty}
    \mathrm{MSE}_{\mathrm{first}}(M,B)
    =[p(B)-b]^2.
    \label{eq:mse_first_limit}
\end{equation}

In the limit $B\to\infty$, both conditional completion probabilities approach unity, so that $q(B)\to1$ and $p(B)\to b$. The exact MSE then reduces to the ideal Bernoulli result
\begin{equation}
    \mathrm{MSE}_{\mathrm{ideal}}(M)
    =\frac{b(1-b)}{M}.
    \label{eq:mse_ideal}
\end{equation}
Thus, when every initial trajectory is followed to absorption, First-Trajectory Filtering becomes ordinary independent sampling and is unbiased regardless of the underlying duration distributions.

The filtering strategy also permits aggregation across independent job submissions. Consider $S$ independent jobs, each employing $M$ workers with the same budget $B$. Pooling their completed initial trajectories is statistically equivalent to applying the filtering procedure to $MS$ independent initial trajectories. The expected number of completed trajectories is $MSq(B)$, and the large-sample MSE becomes
\begin{equation}
    \mathrm{MSE}_{\mathrm{first}}^{(S)}
    \simeq
    [p(B)-b]^2+
    \frac{p(B)[1-p(B)]}{MSq(B)}.
    \label{eq:first_batch_mse}
\end{equation}
Batch aggregation therefore reduces the variance without changing the residual censoring bias. At fixed $B$, increasing the number of independent submissions cannot eliminate this systematic error.

The analytical results establish the statistical properties of First-Trajectory Filtering under finite execution budgets. Whether its reduction in systematic bias compensates for the loss of trajectories generated by continuous sampling depends on the available computational resources. We examine this trade-off in the following subsection.

\subsubsection{Cost-Efficiency and Mean-Squared Error Trade-Offs}
\label{sec:mse_comparison}

The preceding analysis provides an exact expression for the MSE of First-Trajectory Filtering. We now compare its statistical efficiency with that of continuous sampling under identical computational resource constraints. Each of the $M$ independent workers is allocated a computational budget $B=\gamma\bar\tau$, corresponding to a total reserved computational capacity $W_{\mathrm{alloc}}=MB$. Continuous sampling uses this allocation to execute successive trajectories until the deadline, whereas First-Trajectory Filtering retains at most one completed trajectory per worker. The former strategy therefore collects more observations, but its larger sample size may not compensate for the systematic renewal bias.

For continuous sampling, the large-$M$ limit of the observed mean is $p_B=r(B)/m(B)$, where $m(B)$ and $r(B)$ are given by Eqs.~\eqref{eq:renewal_mean} and \eqref{eq:renewal_reward}. Since statistical fluctuations vanish as $M$ increases, the corresponding MSE approaches
\begin{equation}
    \lim_{M\to\infty}
    \mathrm{MSE}_{\mathrm{cont}}(M,B)
    =[p_B-b]^2.
    \label{eq:mse_cont_limit}
\end{equation}
Together with Eq.~\eqref{eq:mse_first_limit}, this result shows that both estimators retain a finite systematic error at fixed execution budget, although their biases exhibit different dependences on $B$.

\begin{figure}[t]
    \centering
    \includegraphics[width=\linewidth]{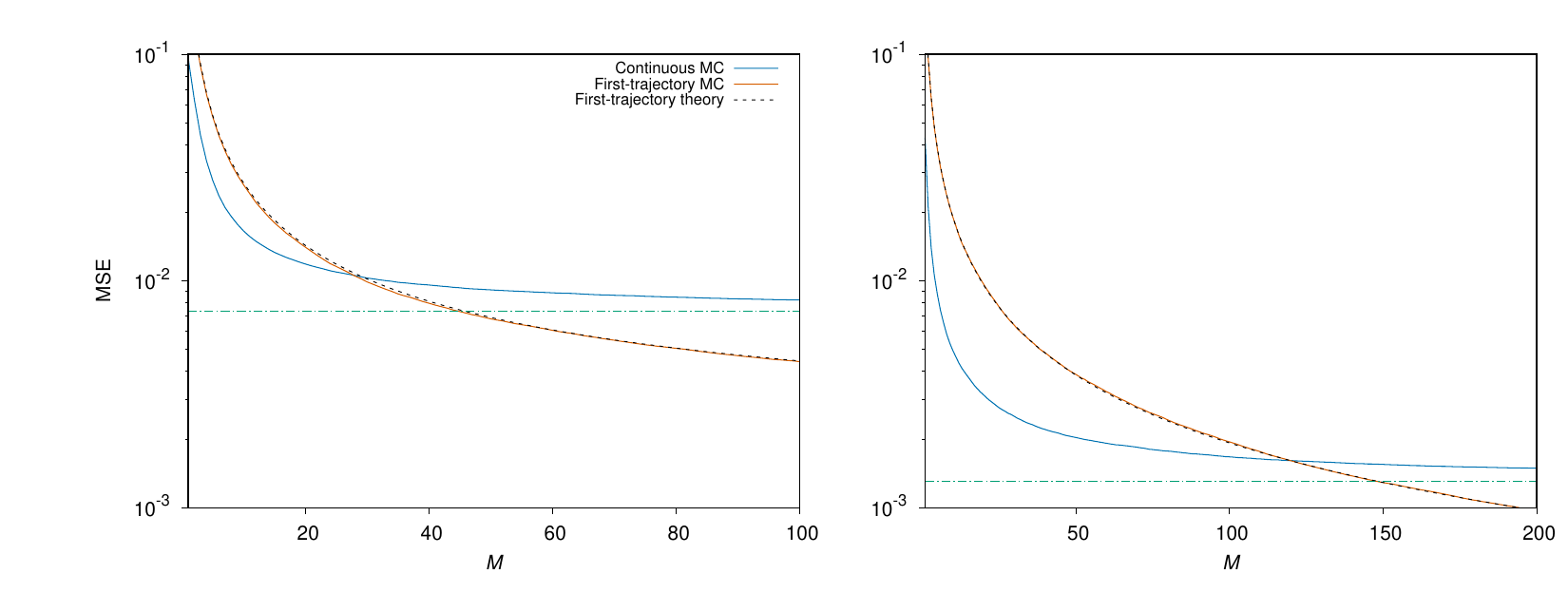}
    \caption{
    Mean-squared error (MSE) as a function of the number of parallel workers $M$ for continuous sampling (blue) and First-Trajectory Filtering (orange), with $b=0.25$, $\tau_1=1/6$, and $\tau_0=5/6$. Results are averaged over $20\,000$ independent Monte Carlo realizations for each $M$. The black dashed curves are the exact analytical predictions for First-Trajectory Filtering, Eq.~\eqref{eq:first_mse_exact}, and the horizontal dash-dotted lines represent the large-$M$ squared bias of continuous sampling, Eq.~\eqref{eq:mse_cont_limit}. Left panel: For $\gamma=2$, First-Trajectory Filtering becomes more accurate than continuous sampling at approximately $M=30$. Right panel: For $\gamma=5$, the crossover shifts to approximately $M=120$. At this larger budget, the First-Trajectory MSE is practically indistinguishable, on the scale shown, from the ideal Bernoulli result $b(1-b)/M$.
    }
    \label{fig:mse_gamma_comparison}
\end{figure}

Figure~\ref{fig:mse_gamma_comparison} compares the MSEs of the two strategies for $b=0.25$, $\tau_1=1/6$, and $\tau_0=5/6$. The results were obtained from $20\,000$ independent Monte Carlo realizations for each worker allocation $M$, using the same exponential renewal process for both estimators. The analytical prediction for First-Trajectory Filtering is given by Eq.~\eqref{eq:first_mse_exact}, while the horizontal lines indicate the large-$M$ limits of continuous sampling obtained from Eq.~\eqref{eq:mse_cont_limit}.

For the shorter budget $\gamma=2$, corresponding to $B=4/3$, continuous sampling initially achieves a smaller MSE because it collects more completed trajectories per worker. As $M$ increases, however, its sampling variance decreases and the systematic renewal bias becomes increasingly important. The continuous-sampling MSE approaches the theoretical floor $[p_B-b]^2\simeq7.32\times10^{-3}$. For First-Trajectory Filtering, the completion probability is $q(B)\simeq0.8485$, and the probability of observing state $X=1$ among completed initial trajectories is $p(B)\simeq0.2945$. The resulting selection bias is approximately $0.0445$. Despite this residual bias and its smaller sample size, First-Trajectory Filtering becomes more accurate at approximately $M=30$, as shown in the left panel of Fig.~\ref{fig:mse_gamma_comparison}.

Increasing the budget to $\gamma=5$, or $B=10/3$, substantially reduces the systematic errors of both strategies. The continuous-sampling MSE now approaches a lower asymptotic floor, $[p_B-b]^2\simeq1.31\times10^{-3}$. Meanwhile, the First-Trajectory completion probability increases to $q(B)\simeq0.9863$, and its selection bias decreases to approximately $0.00348$. The corresponding squared bias, about $1.21\times10^{-5}$, is sufficiently small that the MSE closely follows the ideal Bernoulli behavior $b(1-b)/M$ over the range displayed in the right panel of Fig.~\ref{fig:mse_gamma_comparison}. Because continuous sampling also benefits from the longer execution budget, the crossover shifts to approximately $M=120$.

\begin{comment}
An analytical estimate of the worker allocation required for First-Trajectory Filtering to outperform continuous sampling can be obtained by comparing Eq.~(37) with the large-$M$
squared-bias floor of continuous sampling, Eq.~(41). Defining $\Delta_{\mathrm{cont}}(B)=r(B)/m(B)-b$ and $\Delta_{\mathrm{first}}(B)=p(B)-b$, the intersection occurs approximately at
\begin{equation}
    M_{\mathrm{int}}(B)
    \simeq
    \frac{p(B)[1-p(B)]}
    {q(B)\left[
    \Delta_{\mathrm{cont}}^2(B)
    -\Delta_{\mathrm{first}}^2(B)
    \right]},
    \label{eq:crossover_estimate}
\end{equation}
provided that $\Delta_{\mathrm{cont}}^2(B)> \Delta_{\mathrm{first}}^2(B)$. For the parameters of Fig.~2, this expression gives $M_{\mathrm{int}}\simeq46$ for $\gamma=2$ and
$M_{\mathrm{int}}\simeq148$ for $\gamma=5$. Both values exceed the corresponding crossover allocations observed in the simulations. Thus, comparison with the continuous sampling bias floor provides a useful analytical estimate of the worker scale beyond which filtering becomes advantageous. The estimate is asymptotic, since the finite-$M$ continuous-sampling MSE also contains statistical fluctuations and finite-sample corrections to its mean.
\end{comment}

The agreement between the simulated First-Trajectory MSE and the exact analytical prediction is particularly noteworthy. It confirms that the statistical error can be quantitatively accounted for by the state-dependent completion probabilities and the fluctuations in the number of completed initial trajectories. For continuous sampling, the approach toward the horizontal theoretical limits demonstrates how increasing the number of workers progressively exposes the finite-budget renewal bias.

These results reveal a competition between statistical precision and systematic error. At relatively small worker allocations, continuous sampling can be advantageous because the additional completed trajectories reduce sampling fluctuations. At larger allocations, the variance decreases for both strategies, and the estimator with the smaller systematic bias becomes preferable. The crossover therefore depends on the absorbing-state probabilities, the conditional duration distributions, and the execution budget, rather than defining a universal cluster-size threshold.

From an operational perspective, First-Trajectory Filtering offers a simple means of reducing renewal-induced sampling distortions without modifying the underlying stochastic dynamics. Its main cost is the deliberate rejection of subsequent trajectories, which can make continuous sampling more efficient at modest worker allocations. Figure~\ref{fig:mse_gamma_comparison} quantifies this trade-off and demonstrates that collecting more completed trajectories does not necessarily produce a more accurate estimate of the absorbing-state probability.

\subsubsection{Observability Limits under Severe Censoring}
\label{sec:observability}

The preceding analysis assumes that the absorbing-state probability $b$ and the conditional mean durations $\tau_0$ and $\tau_1$ are known when evaluating the analytical predictions. In practical simulations, however, these quantities are precisely what one seeks to determine. This distinction becomes particularly important under severe censoring, when the computational budget $B=\gamma\bar\tau$ is short compared with the characteristic absorption times.

In absorbing-state dynamics, the final state $X\in\{0,1\}$ is an emergent outcome that becomes observable only when a trajectory reaches absorption. A trajectory interrupted at the execution deadline therefore provides no direct information about its eventual absorbing state. When the budget is small, the completed trajectories constitute a strongly selected subset of the underlying population, preferentially representing the faster absorbing pathway.

For First-Trajectory Filtering, the probability of observing state $X=1$ among completed trajectories is given by Eq.~\eqref{eq:first_probability}. In the strongly censored limit $B\to0^+$, the exponential completion probabilities satisfy
\begin{equation}
    q_x(B)=\frac{B}{\tau_x}+O(B^2),
    \qquad x\in\{0,1\},
\end{equation}
so that
\begin{equation}
    \lim_{B\to0^+}p(B)
    =
    \frac{b/\tau_1}
    {b/\tau_1+(1-b)/\tau_0}.
    \label{eq:severe_censoring_limit}
\end{equation}
Thus, the observed absorbing-state fraction does not approach the true prior $b$ as the execution budget vanishes. Instead, it is controlled by both the absorbing-state probabilities and the relative rates of the competing absorption pathways. For the parameters considered here, Eq.~\eqref{eq:severe_censoring_limit} gives $p(0^+)=0.625$, substantially exceeding the true value $b=0.25$.

The difficulty is compounded by the decreasing number of completed trajectories. Since
\begin{equation}
    q(B)
    =
    B\left(
    \frac{b}{\tau_1}
    +\frac{1-b}{\tau_0}
    \right)+O(B^2),
    \label{eq:small_budget_completion}
\end{equation}
the expected number of completed initial trajectories is
\begin{equation}
    \mathbb{E}[J]=Mq(B)=O(MB).
\end{equation}
At fixed $M$, reducing the execution budget therefore simultaneously increases the selection bias and decreases the amount of information available for estimating the underlying dynamics.

If the conditional duration distributions are known, the completion probabilities $q_x(B)$ can be calculated and used to correct the selection bias, for example through inverse probability weighting (IPW)~\cite{Horvitz_1952,Robins_1994,Seaman_2013}. However, when $b$, $\tau_0$, and $\tau_1$ are unknown, the required correction probabilities must themselves be estimated from the censored observations. In the strongly censored regime, this additional inference problem can substantially increase statistical uncertainty and make the resulting estimates sensitive to assumptions about the unobserved portions of the duration distributions.

For the exponential model, the observable data from an initial trajectory consist of the recorded time $Y=\min(T,B)$, the completion indicator $\delta=\mathbb{I}(T\leq B)$, and the absorbing state $X$ when $\delta=1$. The corresponding state-specific completion densities are
\begin{equation}
    g_x(t)=
    \frac{P(X=x)}{\tau_x}
    e^{-t/\tau_x},
    \qquad 0<t\leq B,
    \label{eq:observed_completion_density}
\end{equation}
while the probability of remaining unabsorbed at the deadline is
\begin{equation}
    S(B)=
    be^{-B/\tau_1}
    +(1-b)e^{-B/\tau_0}.
    \label{eq:censoring_survival}
\end{equation}
These expressions permit likelihood-based inference of the model parameters in principle. Nevertheless, when $B$ is small, the observed completion times cover only a narrow portion of the duration distributions, making the simultaneous estimation of $b$, $\tau_0$, and $\tau_1$ increasingly demanding.

More generally, without assuming a specific parametric form for the conditional duration distributions, observations restricted to $t\leq B$ cannot determine how the probability mass remaining at the deadline is ultimately divided between the two absorbing states. Standard survival and competing-risks methods, including the Kaplan--Meier and Aalen--Johansen estimators~\cite{Kaplan_1958,Aalen_1978}, can characterize event probabilities within the observed time interval, but cannot determine the unrestricted eventual absorption probabilities without additional information about events beyond the observation horizon.

The central operational implication is therefore not that statistical correction is impossible under severe censoring, but that its reliability depends critically on the information available about the absorption-time distributions. When the underlying parameters are unknown, extending the computational budget may be more effective than increasing the number of short, heavily censored trajectories. In particular, First-Trajectory Filtering provides a simple and nearly unbiased estimator only when the budget is sufficiently long that the state-dependent completion probabilities are both close to unity. 

\section{Application: Renewal Bias in the Axelrod Model}
\label{sec:axelrod}

To illustrate that the renewal bias predicted by our minimal model also occurs in a spatially interacting system, we consider the Axelrod model of cultural dissemination~\cite{Axelrod_1997}. The system consists of an $L\times L$ square lattice in which each site carries a cultural vector of $F$ nominal features, each taking one of $q$ possible traits. Neighboring sites interact with a probability proportional to their cultural similarity, defined as the fraction of shared features. Upon interaction, one differing feature is modified to match the corresponding feature of the neighboring site. The dynamics eventually reach an absorbing configuration in which no neighboring pair has partial cultural overlap.

The Axelrod model exhibits a nonequilibrium transition between monocultural and fragmented absorbing configurations as the trait diversity $q$ increases~\cite{Castellano_2000,Klemm_2003,Reia_2016,Campos_2025}. For $F=3$ and $L=200$, the transition occurs between $q=14$ and $q=15$. Near this boundary, two qualitatively distinct classes of absorbing configurations coexist: low-domain-density configurations, characterized by long absorption times, and highly fragmented configurations, which typically freeze much more rapidly. This pronounced difference in their characteristic absorption times is the physical motivation for the state-dependent durations introduced in our minimal renewal model.

\begin{figure}[t]
    \centering
    \includegraphics[width=\linewidth]{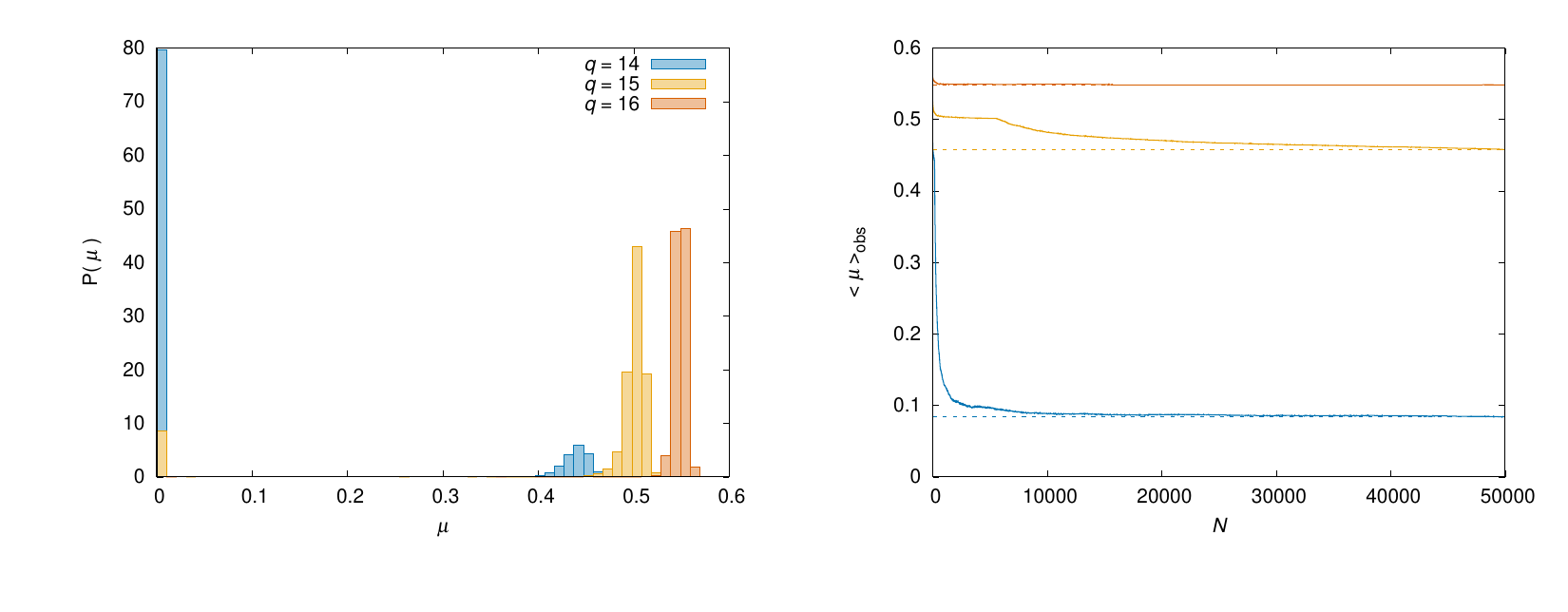}
    \caption{
    Renewal sampling bias in the Axelrod model for $F=3$ and $L=200$.
    Left panel: Probability distributions $P(\mu)$ of the final domain density $\mu$ for $q=14$, $15$, and $16$, illustrating the two distinct classes of absorbing configurations near the nonequilibrium transition.
    Right panel: Cumulative observed mean $\langle\mu\rangle_{\mathrm{obs}}$ obtained from $500$ continuously renewed workers, with sampling terminated after $50\,000$ completed trajectories. Horizontal dashed lines indicate unbiased reference estimates from $5000$ initial trajectories, selected independently of their absorbing outcomes and followed to absorption. The upward displacement of the continuous-sampling estimates reflects the preferential representation of rapidly absorbing, fragmented configurations.
    }
    \label{fig:axelrod_results}
\end{figure}

The left panel of Fig.~\ref{fig:axelrod_results} shows the probability distributions $P(\mu)$ of the final domain density $\mu$ for $q=14$, $15$, and $16$. The distributions reveal two distinct classes of absorbing configurations, associated with low and high domain densities. Their peaks are broadened by finite-size effects, but remain sufficiently separated to motivate a binary description of the absorbing outcomes. The minimal model does not attempt to reproduce the detailed distributions of $\mu$ or the absorption times; rather, it isolates the essential correlation between the nature of the absorbing configuration and the time required to reach it. For $F=2$, the peaks overlap more substantially, making the distinction between the two absorbing classes less pronounced~\cite{Reia_2026}.

The right panel of Fig.~\ref{fig:axelrod_results} illustrates the corresponding sampling distortion. Each of the $500$ workers immediately initiates a new trajectory upon absorption, and the simulation is terminated after a prescribed number of completed trajectories rather than at a common execution deadline. Rapidly freezing fragmented configurations are consequently overrepresented during the transient accumulation of completed trajectories, shifting the observed mean $\langle\mu\rangle_{\mathrm{obs}}$ toward higher domain densities. The reference estimates are obtained from initial trajectories selected independently of their outcomes and followed to absorption, thereby avoiding the selection associated with the continuous renewal protocol. Unlike the fixed-budget exponential model, this experiment uses a completion-count stopping rule; its purpose is to illustrate the same underlying correlation between absorption times and the composition of the observed sample.

These results demonstrate that the sampling mechanism isolated by the minimal exponential model is relevant to a nontrivial spatial system. The Axelrod simulations are not intended as a quantitative test of the exponential duration assumption, but rather as a physical illustration of the underlying mechanism: when distinct absorbing outcomes have markedly different characteristic times, continuous renewal sampling can distort their relative representation. The binary model captures this essential feature without requiring a detailed description of the microscopic dynamics.

\section{Discussion}\label{sec:discussion}

The results presented here demonstrate that asynchronous parallel simulations of absorbing processes can exhibit systematic sampling bias when the time required to reach absorption depends on the eventual absorbing state. By representing independent workers as renewal processes with state-dependent trajectory durations, we have shown that pooling all trajectories completed within a fixed computational budget does not, in general, reproduce the underlying absorbing-state probabilities. Instead, faster absorbing pathways are overrepresented among completed trajectories, producing a systematic distortion that persists as the number of parallel workers increases at fixed execution time.

The exponential model allows this effect to be characterized analytically. In particular, renewal theory provides exact expressions for the expected number of completed trajectories and the corresponding absorbing-state counts, yielding a prediction without adjustable parameters for the macroscopic evolution of the observed mean. The resulting bias decreases algebraically with the execution budget, approaching zero in the long-time limit. Thus, the distortion is not an intrinsic property of the stationary renewal process, but a finite-budget effect associated with the initialization and termination of asynchronous sampling. Increasing the number of workers reduces statistical fluctuations around the biased finite-budget mean, but does not eliminate the bias itself.

First-Trajectory Filtering provides an alternative sampling strategy by retaining only the initial trajectory attempted by each worker. This procedure eliminates the repeated-renewal mechanism responsible for the transient bias of continuous sampling. However, because absorption times are exponentially distributed and therefore unbounded, some initial trajectories remain incomplete at any finite execution deadline. The resulting selection bias favors faster absorbing pathways and prevents the filtered estimator from being exactly unbiased. Nevertheless, its bias decreases exponentially with the execution budget, in contrast to the algebraic decay found for continuous sampling. This difference in convergence rates constitutes a central advantage of First-Trajectory Filtering.

The competition between statistical variance and systematic bias is quantified by the mean-squared error. Continuous sampling generally produces more completed trajectories per worker and can therefore achieve greater statistical precision at small worker allocations. As the number of workers increases, however, sampling fluctuations decrease and the residual bias becomes increasingly important. For the parameters investigated here, the MSE comparison reveals crossovers near $M=30$ and $M=120$ for budget ratios $\gamma=2$ and $\gamma=5$, respectively. These crossovers are not universal thresholds, but depend on the absorbing-state probabilities, the conditional duration distributions, and the available computational budget. The close agreement between the simulations and the analytical First-Trajectory MSE confirms that the trade-off can be understood quantitatively within the renewal framework.

An additional difficulty arises when the absorbing-state probabilities and characteristic durations are unknown, as is typically the case in simulations of complex nonequilibrium systems. Under severe right-censoring, the absorbing states of incomplete trajectories remain unobserved, while the completed sample is strongly enriched in faster pathways. Although statistical corrections are possible when the conditional duration distributions are known, estimating the required completion probabilities from heavily censored data can be inefficient and sensitive to modeling assumptions. For unrestricted duration distributions, observations within a finite time horizon cannot, in general, determine how the remaining probability mass is distributed among eventual absorbing states. Consequently, increasing the number of short simulations cannot necessarily compensate for insufficient observation time.

The application to the Axelrod model~\cite{Axelrod_1997} provides a physical illustration of the mechanism identified by the renewal analysis. Near its nonequilibrium transition~\cite{Castellano_2000,Klemm_2003,Reia_2016,Campos_2025,Reia_2026}, the system exhibits two distinct classes of absorbing configurations with markedly different characteristic absorption times. This separation motivates the binary structure of the minimal model, while the systematic overrepresentation of rapidly absorbing fragmented configurations under continuous execution demonstrates the resulting sampling bias. The independently sampled reference trajectories, all followed to absorption, provide an unbiased reference estimate. Thus, the exponential model is not intended to reproduce the detailed dynamics of the Axelrod system, but to isolate a general statistical mechanism that can arise whenever absorbing outcomes are correlated with their characteristic transient durations.

These findings have practical implications for high-performance simulations of absorbing dynamics. The appropriate sampling strategy depends not only on the number of trajectories collected, but also on the relationship between their characteristic durations and the available execution time. When computational budgets are sufficiently long, continuous sampling approaches unbiased estimation and benefits from its larger sample volume. At shorter budgets and large worker allocations, First-Trajectory Filtering can substantially reduce the systematic error, even though it discards trajectories generated after the initial renewal cycle. If initial trajectories can be allowed to reach absorption without premature termination, the filtered estimator reduces to ordinary independent sampling and is unbiased regardless of the underlying duration distributions.

The central conclusion is that computational throughput alone does not guarantee statistical accuracy in asynchronous simulations of absorbing processes. At fixed execution budgets, increasing parallelism may improve precision while leaving systematic sampling errors unchanged. Reliable estimation therefore requires accounting for the interplay between trajectory durations, censoring, and the sampling protocol, rather than relying exclusively on the accumulation of larger datasets.

\section*{Acknowledgments}

JFF is partially supported by Conselho Nacional de Desenvolvimento Cient\'{\i}fico e Tecnol\'ogico grant number 300200/2026-9.

\end{document}